\documentclass[]{scrartcl}
\usepackage{latexsym}
\usepackage{cuted}
\usepackage{breqn}
\usepackage{amssymb}
\usepackage{makeidx}
\usepackage{dblfloatfix}
\usepackage[section]{placeins}
\usepackage{bm}
\usepackage{makecell}
\PassOptionsToPackage{hyphens}{url}
\usepackage{hyperref}
\usepackage{url}
\usepackage[english]{babel}
\usepackage{amsfonts}
\usepackage{graphicx}
\usepackage{graphics}
\usepackage{float}
\usepackage{authblk}
\usepackage{multirow}
\usepackage{flushend}

\begin{document}
\title{A new laser-ranged satellite
for General Relativity and Space Geodesy
}
\subtitle{III. De Sitter effect and the LARES 2 space experiment.}

\author[1,2]{Ignazio Ciufolini\thanks{ignazio.ciufolini@unisalento.it}}
\author[3]{ Richard Matzner}
\author[4]{Vahe Gurzadyan}
\author[5]{Roger Penrose}

\affil[1]{\footnotesize  Dip. Ingegneria dell'Innovazione, Universit\`a del Salento, Lecce, Italy}
\affil[2]{Museo della fisica e Centro studi e ricerche Enrico Fermi, Rome, Italy}
\affil[3]{Theory Group, University of Texas at Austin, USA}
\affil[4]{Center for Cosmology and Astrophysics, Alikhanian National Laboratory and Yerevan State University, Yerevan, Armenia}
\affil[5]{Mathematical Institute, University of Oxford, UK }

\renewcommand\Authands{ and }

\date{}
\maketitle
\abstract{
In two previous papers we presented the LARES 2 space experiment aimed at a very accurate test of frame-dragging and at other tests of fundamental physics and measurements of space geodesy and geodynamics. We presented the error sources of the LARES 2 experiment, its error budget and Monte Carlo simulations and covariance analyses confirming an accuracy of a few parts in one thousand in the test of frame-dragging. Here we discuss the impact of the orbital perturbation known as de Sitter effect, or geodetic precession, in the error budget of the LARES 2 frame-dragging experiment. We show that the uncertainty in de Sitter effect has a negligible impact in the final error budget because of the very accurate results now available for the test of de Sitter precession and because of its very nature.
The total error budget in the LARES 2 test of frame-dragging remains at the level of about $0.2\%$, as determined in the first two papers of this series.
} 

\section{Introduction to the de Sitter effect}
\label{intro}
The LAGEOS-LARES 2 space experiment is designed to achieve a new, accurate measurement of the General Relativistic frame dragging due to the rotation of the Earth. Analytical estimates, covariance studies, and Monte Carlo simulations concur that the expected error level in this effect is of order $0.2\%$, as shown in previous papers in this series (\cite{bib1,bib1bis}). In addition to the frame-dragging, gravitomagnetic effect,  there is another general relativistic perturbation of an orbiting gyroscope, relative to an asymptotic inertial frame: the de Sitter or geodetic (or geodesic) precession \cite{bib2} (see also \cite{bib3}) This precession is due to the coupling between the velocity of a gyroscope orbiting a central body and the static part of the field (Schwarzschild metric) generated by the central mass.

The de Sitter precession can be derived by Fermi--Walker \cite{bib4} transport  along the worldline of a test-gyroscope. We first consider a spacelike spin four-vector $S^\alpha$ at each point of a timelike curve $x^\alpha(s)$ with tangent vector $u^\alpha$. We thus have: $S^\alpha \, u_\alpha \, = \, 0$. According to special relativistic kinematics and to the medium strong equivalence principle (all the laws of physics are the laws of special relativity in a local inertial frame \cite{bib5,bib6,bib7,bib5bis}), the spin vector $S^\alpha$ obeys Fermi-Walker transport along the curve:
\begin{equation}\label{a}
  S^\alpha \, _{;\beta} \, u^\beta \, = \,u^\alpha \,(\, a^\beta \, S_\beta\,) \equiv \, u^\alpha \, ( \, u^\beta \, _{;\gamma} \,u^\gamma \, S_\beta \, );
\end{equation}
where $a^\beta \, \equiv \, u^\beta \, _{;\gamma} ~ u^\gamma$ is the four-acceleration of the test-gyroscope and the semicolon denotes the covariant derivative. Suppose the timelike curve is a geodesic. (Any test-particle in the gravitational field of a massive body follows a timelike geodesic of the spacetime; a timelike geodesic path--world line--in spacetime's Lorentzian geometry is one that locally maximizes proper time, in analogy with the length-minimizing property of Euclidean straight lines. This is the case for a satellite in free fall, affected only by gravitational forces. A geodesic has $zero$ four-acceleration.) Then we have: $u^\beta \, _{;\gamma} ~ u^\gamma \, = \, 0$ and therefore: $S^\alpha \, _{;\beta} \, u^\beta \, = \, 0$. In this case the Fermi--Walker transport is just the parallel transport along the geodesic. Hence the orbital angular momentum of a satellite around the Earth is parallel-propagated as the Earth-satellite system orbits the Sun.

The spacetime metric generated by a matter distribution with rest-mass density $\rho$ can be written, in a weak gravitational field, at the order of approximation beyond Newtonian theory (the post-Newtonian order) in terms of the standard Newtonian potential $U$, solution of $\Delta \, U \, = \, -4 \, \pi \, \rho$. In isotropic coordinates, we have \cite{bib7,bib9}:
\begin{equation}\label{b}
ds^2 = - (1 - 2 U + 2 \beta U^2 ) dt^2 + (1 + 2 \gamma U ) \delta_{i k} dx^i dx^k
\end{equation}

\noindent where $\beta$ and $\gamma$ are the two main post--Newtonian parameters, both equal to 1 in general relativity, representing respectively the non-linarity in the superposition law for gravity of the gravitational interaction (interpretation valid only in the so-called standard PPN gauge) and the amount of space curvature produced by a mass \cite{bib7,bib9}. For simplicity, in expression (2), we have set equal to zero the other PPN parameters characteristic of preferred--frame, preferred--location and non--conservative theories. For $\beta$ and $\gamma$ equal to 1, in standard non-isotropic coordinates, the metric is just the post--Newtonian approximation of the Schwarzschild metric \cite{bib6,bib7}.

To analyse the behaviour of the orientation of the spin of a test--gyroscope with respect to an asymptotic inertial frame, i.e. with respect to the distant stars (assuming that the universe is not rotating \cite{bib7}), we introduce, at each point along the timelike world--line of the test-gyroscope, a local orthonormal frame \cite{bib11}:
 $ { \bm{\lambda}\/}_{(\mu)}$:
 $ g_{\alpha \beta}  \lambda^\alpha_{(\nu)}$ $\lambda^\beta_{(\mu)}=$ $\eta_{\nu \mu}$. The index between parentheses:
 $\mu \, \epsilon \, ( \, 0, \, 1, \, 2, \, 3 \,)$ is the label of each vector of the local tetrad, or local vierbein (the indices between parentheses can be raised and lowered using
 $\eta^{\alpha \beta}$ and $\eta_{\alpha \beta}$, i.e. the Minkowski metric, and as usual the standard indices using $g^{\alpha \beta}$ and $g_{\alpha \beta}$.)
By construction the timelike vector of this tetrad is the four--velocity of the test-gyroscope: $\lambda^\alpha \, _{(0)}  \equiv  u^\alpha$ and the spatial
 axes of the tetrad are non--rotating with respect to an asymptotic frame where $g_{\alpha \beta} \, \rightarrow \, \eta_{\alpha \beta}$. In other words the
 spatial axes are non-rotating with respect to a local orthonormal tetrad at rest in the asymptotically flat coordinate system of the metric (2). Therefore, the
 spatial axes of the tetrad are not Fermi—Walker transported along the timelike world line of the test-gyroscope, but are obtained at each point with pure
 Lorentz boosts--with no spatial rotations--between a local frame at rest in the coordinate system of the asymptotically flat, post-Newtonian, metric (2) and
 an observer moving with the test-gyroscope along the curve $x^\alpha(s)$ with four--velocity $u^\alpha$. The spatial axes $\bm{ \lambda} \,_{(i)}$
 of this local frame may be thought of as physically realized by three orthonormal telescopes, always pointing towards the same distant stars fixed with
 respect to the asymptotic inertial frame where: $g_{\alpha \beta} \longrightarrow \eta_{\alpha \beta}$ (negecting abberation, which produces periodic
 variations in the pointing direction, but which averages out over one orbit). The vector $S^\alpha$ may be thought of as physically realized by the spacelike
 angular momentum vector of a spinning particle or test-gyroscope. Since $S^\alpha$ is a spacelike vector: $S^\alpha \, u_\alpha$ = 0, in the local frame 
${\bm{\lambda}\/}_{(\alpha)}$ we get :$S^{(0)}$ = 0.
After some calculations, we finally get, at the post-Newtonian order.

\begin{eqnarray}
 \nonumber
  \frac{dS_{(i)}}{ds} &=& \epsilon_{(i)(j)(k)} \dot{\Omega}_{(j)}S_{(k)}\\
  \dot{\Omega}_{(j)}  &=& \epsilon_{(j)(l)(m)}\bigg( -\frac{1}{2}v_{l}a_{m}+\big(  \frac{1}{2} +\gamma \big) v_{l}U_{,m} \bigg)
\end{eqnarray}

\noindent where $\epsilon_{(i) \, (j) \, (k)}$ is the Levi-Civita pseudotensor and the comma $_{,m}$ denotes the standard partial derivative with respect to \textit{x}$^{m}$ and in standard vector notation:

\begin{eqnarray}
 \nonumber 
  \frac{d\bm{S}}{ds} &\equiv&  \dot{\bm{\Omega}} \times S\\
  \dot{\bm{\Omega}}  &=& -\frac{1}{2}\bm{v} \times \bm{a} +\big(  \frac{1}{2} +\gamma \big) \bm{v} \times \bm{\nabla}U
\end{eqnarray}
where ${\bm{ S}\/} \, \equiv \left ( \, S_\alpha \, \lambda \, ^\alpha _{(i)} \right )$ and $\bm{\nabla}$ is the standard gradient operator.

This is the precession of a gyroscope with respect to an asymptotic inertial frame in the spacetime metric generated by a static distribution of matter. The first term of (4) is the Thomas precession, due to the non--commutativity of non--aligned Lorentz transformations and to the non--gravitational acceleration \textit{\textbf{a}}. The second term is the de Sitter or geodetic precession \cite{bib2} (see also \cite{bib3}), which may be interpreted \cite{bib6,bib7,bib9} as due to a part (contributing with $\frac{1} {2}$) analogous to the Thomas precession, arising from the non--commutativity of non--aligned Lorentz transformations of special relativity and to the gravitational acceleration ${\bm{ \nabla}} \, U$, that is due to Fermi-Walker transport and to the gravitational acceleration (that might be derived even in the flat spacetime of special relativity), plus a part (contributing with $\gamma$) due to Fermi-Walker transport and to the space curvature of General Relativity measured by the $\gamma$ parameter. In other words the de Sitter precession is the sum of two parts. One part, with factor $\frac{1}{2}$, is due to the time-time component of the metric tensor: $g_{00}$ (in standard post--Newtonian coordinates). If one writes: $g_{00} \, \simeq \, - \, 1 \, + \, 2 \, U \, - \, 2 \, \beta \, U^2$, this effect is due to the second term in $g_{00}$. The other part, parametrized by $\gamma$ (equal to 1 in General Relativity), is due to the spatial curvature measured by $\gamma$ in the space-space components of the metric $g_{ij} \, \simeq \, \left ( \, 1 \, + \, 2 \, \gamma \, U \, \right) \, \delta_{ij}$. This effect was discovered in 1916 by de Sitter \cite{bib2} (see also \cite{bib3})

Thus, in the weak field, slow motion limit, the de Sitter precession of a gyroscope orbiting a central mass $M$, where $U = M/r$ is given by:

\begin{equation}\label{eq12}
\dot {\bm {\Omega}}^{de \, Sitter} \, = \, - \, ({\frac{1}{2} + \gamma}) \,
{\bm{v}} \, \times \, {\bm{ r}} \, \frac{M}{r^{3}} \,  \, \cong \,( - \,19.2 \, milliarcsec/year) ~ {\bm{ \hat{ n}}}
\end{equation}

\noindent where ${\bm{v}}$ is the velocity of the orbiting gyroscope, ${\bm{ r}}$ the distance from the central mass to the gyroscope and $M$ the mass of the central body as measured in the weak field region. For the Earth -- satellite gyroscope orbiting the Sun, this precession is about an axis perpendicular to the ecliptic plane: ${\bm {\hat n}}$. The orbital plane of a satellite orbiting the Earth, such as LARES 2, LAGEOS or the Moon, may be thought of as a huge gyroscope orbiting the Sun and is thus affected by the solar de Sitter precession. Thus the solar de Sitter precession changes the nodal longitude of any satellite orbiting the Earth, measured relative to an asymptotic inertial frame, by:

\begin{equation}\label{eq21}
19.2\,  milliarcsec/year \, \times \, \cos \, 23.5^{o} \,\cong \, 17.6 \, milliarsec/year
\end{equation}
where $23.5^{o}$ is the obliquity of the ecliptic.

The de Sitter effect on a gyroscope due to the mass of the Sun has been accurately measured on the Moon-Earth ``gyroscope'' by Lunar Laser Ranging (see next section). The de Sitter effect on a gyroscope orbiting the Earth, due to the mass of the Earth, was measured by the Gravity Probe B experiment, but the Earth's de Sitter effect does not directly affect the LAGEOS -- LARES 2 observation.

\section{Nature of the de Sitter effect and its impact in the accuracy of the LARES 2 experiment}

The de Sitter precession is a simple consequence of the gravitational field generated by a static, non-rotating, mass and has been measured, in weak field, by a huge number of highly accurate experiments (see, e.g.\cite{bib7,bib9,bib11tris,bib19,bib12}). On the other hand, the frame-dragging effect is a consequence of the gravitational field of a rotating mass (such as a spinning black hole or the spinning Earth). For a discussion on the interpretation of the Lunar Laser Ranging observation of frame dragging, see, e.g. \cite{tredicibis}. The only tests of the spacetime solution of Einstein's General Relativity generated by a rotating mass (which has a key role in a number of astrophysical processes such as the emission of gravitational waves by a system of two coalescing spinning black holes) have been the LAGEOS plus LAGEOS 2, GP-B, and LARES tests, which have moderate accuracy. The LARES test is so far the most accurate test of frame-dragging with an uncertainty of about 5\%, and is aimed to reach an uncertainty of about 2\%. The LARES 2 experiment will improve the tests of frame-dragging to reduce the total error to a few parts in a thousand.

The de Sitter precession adds to the frame-dragging, but there exists no viable theory of
gravitation that is both in agreement with all the existing tests of gravitational physics, and that predicts a solar system violation of the de Sitter precession at a level larger than $8.7 \times 10^{-6}$ (about 9 parts per million) of the value predicted by General Relativity, negligible at the level of the frame-dragging accuracy.

The frame-dragging is independent of the de Sitter precession: there are indeed viable gravitational theories (such as the Chern-Simons theory and other f(R) theories (\cite{bib13}) which agree with all the gravitational tests except frame-dragging. The Chern-Simons gravitational theory is equivalent to heterotic string theories of type II with relevant implications for the explanation of one of the biggest riddles of physics today, the nature of dark-energy and quintessence, a mysterious form of energy calculated to constitute 70\% of our universe. These theories predict a different outcome than General Relativity for frame-dragging around a rotating body, such as the Earth or a spinning black hole (\cite{bib13}).

In the slow motion weak field situation appropriate for satellite motion near the Earth, both the Earth's dimensionless gravitational potential and the Sun's are small, of order $10^{-8}$ or less. Then the post Newtonian approximation holds for that satellite motion and the de Sitter precession, as derived in the previous section, can be written as the simple expression Eq. (5). As discussed in section 1, the first term $\frac{1}{2}$ in Eq. (5) can be interpreted as the Thomas precession, an essentially Special Relativity effect under the central gravitational acceleration toward the mass $M$. The second term is proportional to the post Newtonian parameter $\gamma$ which describes the spatial curvature generated by the non-rotating central mass.  Thus the deSitter precession is a simple consequence of Special Relativity, of the extremely well tested equivalence principle (freely falling test particles follow timelike geodesics) and of the space curvature generated by a static mass, as parametrized by $\gamma$.

In the solar system, $\gamma$ has been measured by a large number of experiments. The most accurate determination of $\gamma$ so far is its measurement by the CASSINI spacecraft in the gravitational field of the Sun (\cite{bib15}), with an accuracy of about $2.3 \times 10^{-5}$ (we emphasize that the orbital plane if the LARES 2 satellite will be affected by the de Sitter precession due to the mass of the {\itshape Sun}.) From Eq. (5) this implies fractional error in the de Sitter precession of $\frac{2}{3}$ the fractional error in $\gamma$, yielding a fractional error about  $1.53 \times 10^{-5}$ in the value of the de Sitter precession in the field of the Sun. The de Sitter precession perturbs the node of the LAGEOS satellite with a shift of $17.6$ $milliarcsec/yr$ (see the previous section and \cite{bib16}). Therefore the uncertainty in the de Sitter precession on the LAGEOS node is about $ 1.53 \times 10^{-5} \times 17.6$ $milliarcsec/year = 2.7 \times 10^{-4}$ $milliarcsec/yr$. The total frame-dragging effect on the node of LAGEOS is about $31$ $milliarcsec/yr$ (as first calculated in \cite{bib17,bib18}), giving a fractional error in the frame dragging at a level of $(2.7 \times 10^{-4}$ $milliarcsec/yr)/(31$  $milliarcsec/yr)$ $ = 8.7 \times 10^{-6}$, which is negligible in the LARES 2-LAGEOS measurement of frame-dragging.

\section{Direct tests of the de Sitter effect}

Instead of considering the tests of the space curvature parameter $\gamma$ one could consider only experiments {\itshape directly} measuring the de Sitter precession. The solar de Sitter precession was directly measured by the Lunar Laser Ranging (LLR) experiment, with an accuracy of about $5 \times 10^{-3}$  \cite{bib12}.
 Translated to frame dragging, this error by itself amounts to $(5 \cdot 10^{-3}  \times 17.6$ $milliarcsec/yr) / (31$ $milliarcsec/yr) = 0.28\%$ error. However the group of the Leibniz Universit\"at Hannover has recently announced \cite{bib20,bib20bis} that they have reanalysed the LLR data finding a value of  the lunar geodetic precession measured to an accuracy of $1 \times 10^{-3}$ . This gives a $0.06\%$ contribution to the LARES 2 frame dragging determination, still keeping the total (RSS) error budget at the level of about $0.2\%$.

In conclusion, the results summarized in Table 1 show that the error due to the uncertainty in de Sitter effect is negligible in the measurement of frame-dragging with LARES 2.

\begin{table}[h]
  \centering
  \resizebox{\columnwidth}{!}{\begin{tabular}{|l|l|l|}
     \hline
      & \makecell[l]{Error in the LARES 2 test\\ of frame-dragging due to\\ the uncertainty in the\\ de Sitter precession only} & \makecell[l]{Total error budget (e.b.)\\ in the LARES 2 test\\ of frame-dragging} \\ \hline
     \makecell[l]{a) Assuming the tests of \\space curvature due to\\ the Sun static mass ($\gamma$ \\accuracy by CASSINI:\\ 2.3 $\times$ 10$^{-5}$ \cite{bib15})}  & \makecell[l]{8.7 $\times$ 10$^{-6}$ error\\ in the LARES 2 test of\\ frame-dragging}& $\approx$ 0.2\% total e.b.\\ \hline
      \makecell[l]{b) Assuming only the\\ tests of de Sitter\\ precession on the Moon\\ with accuracy:\\ 1 $\times$ 10$^{-3} \cite{bib20,bib20bis}$ }  & \makecell[l]{ 0.06\% error\\ in the LARES 2 test of\\ frame-dragging} & $\approx$ 0.2\% total e.b. \\
     \hline
   \end{tabular}}
  \caption{The error introduced by the de Sitter precession in the test of frame-dragging using LARES 2 and its impact in the total error budget.}\label{tab1}
\end{table}

\section{Summary and Conclusions}

To summarize the situation with respect to the error in knowledge of the de Sitter precession: a) if we consider the solar system tests of General Relativity, mainly testing the gravitational field generated
by the static, non-rotating, mass of the Sun, then the test of the space curvature parameter $\gamma$ (known with an uncertainty of about 2.3$\cdot$ 10$^{-5}$) implies a fractional uncertainty of about 8.7
$\cdot$ 10$^{-6}$ in the test of frame-dragging by the LARES 2 experiment; b) if one insists on considering only the {\itshape direct} tests of the de Sitter precession on Earth satellites by the Sun, then the recent results of reanalysis of the LLR data have an uncertainty of 1 $\times 10^{-3}$ in the measurement of the de Sitter precession which implies a fractional uncertainty of about $0.06\%$ in the test of frame-dragging by the LARES 2 experiment. In both cases, as summarized in Table 1, the error introduced by the uncertainty in de Sitter precession is negligible in the measurement of frame-dragging with LARES 2 and the total error budget in the LARES 2 test of frame-dragging remains at the level of about $0.2\%$, as found in earlier papers in this series.

\section{Acknowledgements}
We gratefully acknowledge the Italian Space Agency for the support of the LARES and LARES 2 space missions under agreements No. 2017-23-H.0 and No. 2015-021-R.0. We are also grateful to the International Ranging Service, ESA, AVIO and ELV. RM acknowledges NASA Grant NNX09AU86G and NSF Grant PHY-1620610.


\begin{thebibliography}{}

\bibitem{bib1}
I. Ciufolini et al., A new laser-ranged satellite for General Relativity and Space Geodesy. I. Introduction to the LARES 2 space experiment,  \textit{Eur. Phys. J. Plus} \textbf{132}, 336 (2017).
\bibitem{bib1bis}
I. Ciufolini et al., A new laser-ranged satellite for General Relativity and Space Geodesy. II. Monte Carlo Simulations and covariance analyses of the LARES 2 Experiment, \textit{Eur. Phys. J. Plus} \textbf{132}, 337 (2017).
\bibitem{bib2}
W. de Sitter, On Einstein's theory of gravitation and its astronomical consequences,  Mon. Not. R. Astron. Soc., {\bfseries 77}, 155--184 and 481 (1916), 155--184 and 481.
\bibitem{bib3}
A. Fokker, The Geodesic Precession; a Consequence of Einstein's Theory of Gravitation,  Proc. K. Ned. Akad. Wet., {\bfseries 23.5},  (1921), 729--738.
\bibitem{bib4}
E. Fermi, Sopra i Fenomeni che Avvengono in Vicinanza di una Linea Oraria, Atti Accad. Naz. Lincei Rend. Cl. Sci. Fis. Mat. Nat {\bfseries 31},  (1922), 21--23, 51--52, 101--103.
\bibitem{bib5}
S. Weinberg, Gravitation and Cosmology: Principles and Applications of the General Theory of Relativity (Wiley, New York, 1972).
\bibitem{bib6}
C.W. Misner, K. S. Thorne, and J. A. Wheeler, Gravitation, Freeman, San Francisco, (1973).
\bibitem{bib7}
I. Ciufolini and J. A. Wheeler, Gravitation and Inertia (Princeton Univ. Press, 1995)
\bibitem{bib5bis}
I. Ciufolini, Dragging of Inertial Frames, Nature \textbf{449}, 41-47 (2007).
\bibitem{bib9}
C.M. Will, Theory and Experiment in Gravitational Physics, 2nd edition (Cambridge University Press, Cambridge, UK, 1993).
\bibitem{bib11}
J. L. Synge, Relativity the General Theory, 5th edition (North Holland, Amsterdam 1976).
\bibitem{bib11tris}
S. Turyshev, Experimental Tests of General Relativity: Recent Progress and Future Directions, Physics-Uspekhi, \textbf{52}, 1-27 (2009).
\bibitem{bib19}
 S. Kopeikin, M. Efroimsky, G. Kaplan, Relativistic Celestial Mechanics of the Solar System (Wiley Online Library, 2011).
\bibitem{bib12}
J. M\"{u}ller, L. Biskupek, F. Hofmann and E. Mai, Lunar Laser Ranging and Relativity, Frontiers in Relativistic Celestial Mechanics - Volume 2: Applications and Experiments, ed. S. Kopeikin, 103-156, Berlin, Boston: deGruyter (2014).
\bibitem{tredicibis}
 S. Kopeikin, Y. Xie, Celest. Mech. Dyn. Astron. \textbf{108}, 245 (2010).
\bibitem{bib13}
L. Smith, A. L. Erickcek, R. R. Caldwell and M. Kamionkowski, Effect of Chern-Simons gravity on bodies orbiting the Earth, Phys. Rev. D \textbf{77} (2008), 024015.
\bibitem{bib15}
B. Bertotti, L. Iess and P. Tortora, A test of general relativity using radio links with the Cassini spacecraft, Nature \textbf{425}, 374-376 (2003).
\bibitem{bib16}
I. Ciufolini, A Comprehensive Introduction to the Lageos Gravitomagnetic Experiment, from the Importance of the Gravitomagnetic Field in Physics to a Preliminary Error Budget, Int. J. Mod. Phys. A \textbf{4},  (1989), 3083-3145.
\bibitem{bib17}
I. Ciufolini, Theory and Experiments in General Relativity and other Metric Theories, PhD Dissertation, University of Texas, Austin (Ann Arbor Publishers, Michigan, 1984).
\bibitem{bib18}
I. Ciufolini, Measurement of the Lense-Thirring Effect on Lageos and Another High Altitude Laser Ranged Artificial Satellite, Phys. Rev. Lett. \textbf{56},  (1986), 278-281.
\bibitem{bib20}
F. Hofmann, Lunar Laser Ranging - verbesserte Modellierung der Monddynamik und Schätzung relativistischer Parameter, PhD thesis, Leibniz Universit\"at Hannover, (2017).
\bibitem{bib20bis}
J. M\"{u}ller, private communication,(2017).

\end{thebibliography}
\end{document}